\documentclass[runningheads]{llncs}
\usepackage[T1]{fontenc}
\usepackage{graphicx,verbatim}
\usepackage{hyperref}
\usepackage{color}

\usepackage{amssymb}
\usepackage{amsmath}
\usepackage{mathtools}
\usepackage{upgreek}
\usepackage{array, xcolor, colortbl}
\usepackage{multirow}
\usepackage{booktabs}

\definecolor{Gray}{gray}{0.85}

\begin{document}
\newcommand{\verify}[1]{\textcolor{red}{#1}}
\newcommand{\bh}{\mathbf{h}}
\newcommand{\bx}{\mathbf{x}}
\newcommand{\by}{\mathbf{y}}
\newcommand{\bu}{\mathbf{u}}
\newcommand{\bv}{\mathbf{v}}
\newcommand{\bz}{\mathbf{z}}
\newcommand{\un}{\mathds{1}}
\newcommand{\calL}{\mathcal{L}}
\newcommand{\calN}{\mathcal{N}}
\newcommand{\EAc}[1]{\textcolor{cyan}{#1}}
\newcommand{\EA}[1]{\textcolor{magenta}{#1}}

\title{CNVAE: Anatomically Correct Segmentation via Conditional Generative Modeling}
\titlerunning{CNVAE for Anatomically Correct Segmentation}
%
\author{Freddy Jiang\inst{1} \and
Loïc Le Folgoc\inst{2} \and
Elsa Angelini\inst{1,2}
}
\authorrunning{F. Jiang et al.}
%
\institute{Imperial College London, United Kingdom \and
LTCI, Télécom Paris, Institut Polytechnique de Paris, Palaiseau, France
}

\maketitle              
\begin{abstract}

 Supervised deep learning segmentation models, although unparalleled in performance, are not immune to anatomical mistakes. We develop a segmentation-by-generation paradigm extending the Nouveau VAE (NVAE) generative framework for anatomically correct cardiac segmentation. In particular, we propose a novel conditional NVAE model (CNVAE) for conditional shape generation, and explore its capabilities for image segmentation in two scenarios: (1) direct image-to-mask CNVAE for one-step segmentation; (2) mask-to-mask CNVAE for refinement of segmentation masks produced by third party segmentation models. Both models are lightweight and return consistently anatomically valid segmentation masks ($\geq 98\%$ success rate, an improvement of 15\% above state-of-the-art segmentation models) on the cardiac segmentation/editing tasks of interest, without loss of accuracy in segmented contours.

\keywords{Segmentation  \and Generative Model \and Cardiac MRI \and Shape}

\end{abstract}

\section{Introduction}

Supervised deep learning segmentation models~\cite{ronneberger_olafand_fischer_u-net_2015,hatamizadeh_swin_2022} have pushed forward the state-of-the-art across many applications of medical image segmentation. However, state-of-the-art architectures remain prone to anatomical mistakes. Such mistakes can be a barrier to adoption of deep learning solutions and damage their credibility in the eyes of clinicians.

Various types of information can be incorporated as prior knowledge: 
shape constraints~\cite{oktay_anatomically_2018}, topology specifications~\cite{keshwani2020topnet}, adjacency rules between regions~\cite{gupta2022learning}. 
Many ad-hoc solutions to enforce anatomical validity have been designed for specific applications: cardiac segmentation~\cite{bernard_deep_2018}, vessel trees segmentation~\cite{yang2024benchmarking}, airways segmentation~\cite{zheng2021wignsnet}. Most of these anatomy-aided architectures \cite{liu2021anatomy} remain specialized to a given task (image modality and organ), require large sets of annotated data, and/or are not trained end-to-end.

Recent generative models such as hierarchical VAEs~\cite{vahdat_nvae_2020} 
and diffusion models 
have demonstrated their ability to generate highly-realistic images across a range of applications~\cite{ribeiro_high_2023,biffi_explainable_2020,xu_medsyn_2024}. We posit that such generative capabilities can form the basis for a more general solution to anatomical correctness. In this spirit, we propose to recast anatomically valid segmentation as a problem of conditional mask generation, conditioned on the input image. We make the following contributions:
\begin{itemize}
    \item We propose \textbf{a lightweight conditional NVAE model} ($\texttt{CNVAE}_{\text{Seg}}$) that adapts the Nouveau VAE~\cite{vahdat_nvae_2020} framework, designed for unconditional image generation, to conditional image/mask generation \textit{e.g.}, image segmentation.
    \item We adapt our \texttt{CNVAE} model to do \textbf{mask editing}, by correcting anatomically invalid segmentations produced by arbitrary 
    segmentation models, to anatomically valid segmentation masks. We name this model $\texttt{CNVAE}_{\text{Edit}}$.
    \item We apply $\texttt{CNVAE}_{\text{Seg}}$ and $\texttt{CNVAE}_{\text{Edit}}$ to a task of \textbf{cardiac segmentation}, where we show anatomical validity rates between $98\%$-$99\%$.
\end{itemize}

\section{Related Work} 

Prior to the deep learning era, deformable models  
used to incorporate strong anatomical priors, sometimes presented within Bayesian frameworks~\cite{patenaude_bayesian_2011}; later work \textit{e.g.},~\cite{dalca_anatomical_2018}, extends the Bayesian approach in the deep learning era. Nowadays, the dominant U-Net paradigm~\cite{ronneberger_olafand_fischer_u-net_2015} predicts segmentations from a black-box
combination of image-based and shape-based contextual cues learned from training data, which can
fail and lead to anatomical inconsistencies in images or regions where image cues are ambiguous or
absent (\textit{e.g.},~\cite{bernard_deep_2018}). This has generally been remedied by a variety of application-specific strategies. 

More general-purpose strategies include the use of shape-aware losses~\cite{boussaid_shape_2021} that better encapsulate anatomical information than standard pixel-wise losses. For instance,~\cite{oktay_anatomically_2018} trains an autoencoder to encode anatomical shape variations.  
The encoding of anatomical shapes is then used in an MSE loss to penalise deviations in latent space between ground-truth and generated segmentation masks. Adversarial loss terms have also been pursued using this type of shape encoding~\cite{boutillon2022multi}. 
Other losses attempt to preserve the topology of the organ (\textit{e.g.},~via the use of Betti numbers~\cite{berger_topologically_2024}), but can be more cumbersome to implement.

Instead of training a segmentation model to produce anatomically valid masks,~\cite{painchaud_cardiac_2020} adopts a two-step mask editing strategy, whereby (1)~arbitrary segmentation models generate potentially invalid segmentation masks, then (2)~these invalid segmentations are refined into valid masks by manipulating codes in the latent space of a VAE. 
A limitation of the framework is the inability of non-hierarchical VAEs to capture fine-grained geometrical patterns. In that respect, hierarchical VAEs could realize a suitable compromise between anatomical validity and preservation of the fine geometrical features of the segmentation.

Among hierarchical VAEs, Ladder VAEs  
have been applied to the unconditional generation task of cardiac shape modeling~\cite{biffi_explainable_2020}. NVAE~\cite{vahdat_nvae_2020} was originally proposed for realistic face generation and has not been adapted to conditional medical image generation tasks. Related but distinct from our work are~\cite{ribeiro_high_2023} for generation conditioned on non-image based attributes, and~\cite{harvey_conditional_2021} for a conditional generation framework applied to image inpainting. Finally, diffusion models have also been applied to conditional generation tasks such as segmentation~\cite{wu_medsegdiff_2024}.

\section{Methods}

\begin{figure}
\centering\includegraphics[width=0.9\linewidth]{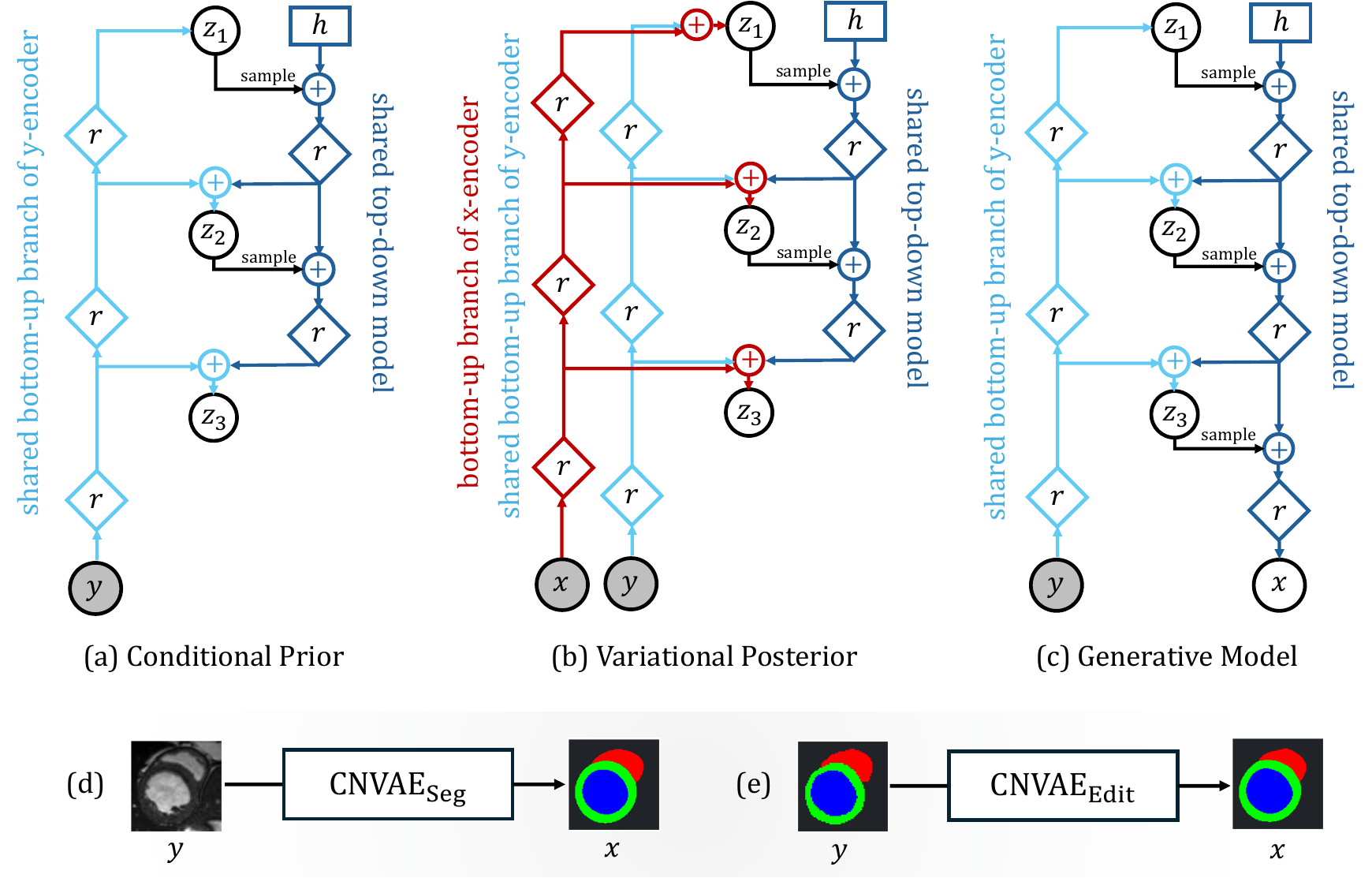}
\caption{CNVAE generative model, with its bidirectional (bottom-up, top-down) hierarchical encoders and decoder, illustrated for a $3$-group hierarchy. $r$ denotes residual neural networks, $+$ denotes feature concatenation, and $h$ is a learnable constant. (a) Conditional prior $p(\bx|\by)$. (b) Variational posterior $q(\bz|\bx,\by)$. (c) Conditional generative model $p(\bx,\bz|\by)$. Based on CNVAE, we propose (d) image-to-mask direct segmentation model; and (e) mask-to-mask editing model.} \label{fig: CNVAE model}
\end{figure}

\subsubsection{Background: NVAE.} Nouveau VAE~\cite{vahdat_nvae_2020} is a hierarchical VAE for image reconstruction and generation. In our case we use it to encode and generate binary shape masks $\bx\in \{0,1\}^{H\times W\times C}$. VAE generative models optimize a joint distribution of the form $p(\bx,\bz)\triangleq p(\bx|\bz)p(\bz)$, where $p(\bz)$ is a prior distribution over the learned latent variables $\bz$, and $p(\bx|\bz)$ is the likelihood function (decoder). Hierarchical VAE learns and manipulates latent variables $\bz \triangleq \{\bz_1,\bz_2, \cdots, \bz_L \}$ that form a hierarchy over $L$ disjoint groups. According to the hierarchy, the prior and the variational posterior factorize as: $p(\bz)\triangleq \prod_l p(\bz_l|\bz_{<l})$ and  $q(\bz|\bx)\triangleq \prod_l q(\bz_l|\bx,\bz_{<l})$. Each factor in these products is modeled as a Normal distribution. VAE models are trained by maximizing an evidence lower bound (ELBO) $\calL_{\text{VAE}}(\bx)$ of the marginal likelihood $\log{p(\bx)}$ over training data $\bx$. The NVAE introduced novel network architecture modules and parameterization of approximate posteriors to provide a computationally-tractable form of hierarchical VAE while scaling up the training to large image sizes~\cite{vahdat_nvae_2020}. 

As we move down in the hierarchy ($l\coloneqq 1\cdots L$), $\bz_l\in \mathbb{R}^{H_l\times W_l\times D}$ consists in an increasingly high-resolution stack of stochastic feature maps ($H_l, W_l \coloneqq H/2^{n_l},W/2^{n_l}$, $n_1>\cdots>n_L$). The variational posterior $q(\bz|\bx)$ is implemented as a bidirectional, bottom-up and top-down Encoder, while the generative model $p(\bx,\bz)$ is implemented as a top-down Decoder with shared parameters with the Encoder.

\subsubsection{Proposed CNVAE.} We extend the NVAE framework to a conditional generative model (conditional NVAE = CNVAE) $p(\bx,\bz|\by)\triangleq p(\bx|\bz,\by)p(\bz|\by)$ for generation of a segmentation mask $\bx$ conditioned on an additional image input $\by\in \mathbb{R}^{H\times W\times C'}$. 
The conditional prior $p(\bz|\by)\triangleq \prod_l p(\bz_l|\bz_{<l},\by)$ and the variational posterior $q(\bz|\bx,\by)\triangleq \prod_l q(\bz_l|\bx,\bz_{<l},\by)$ factorize according to the hierarchy of latent variables. The CNVAE is now trained on pairs of samples $\{\by,\bx\}$ by maximizing an evidence lower bound (ELBO) $\calL_{\text{CVAE}}(\bx,\by)$ of the conditional marginal likelihood $\log{p(\bx|\by)}$, which decomposes following the hierarchy of variables as follows (Eq.~\eqref{eq: ELBO CVAE}):
\begin{multline}
    \calL_{\text{CVAE}}(\bx,\by)=\mathbb{E}_{q(\bz|\bx,\by)}[\log{p(\bx|\bz,\by)}]- \text{KL}[q(\bz_1|\bx,\by)||p(\bz_1|\by)] \\
    - \sum_{l=2}^L \mathbb{E}_{q(\bz_{<l}|\bx,\by)}\big[ \text{KL}[q(\bz_l|\bz_{<l},\bx,\by)||p(\bz_l|\bz_{<l},\by)]\big]
    \label{eq: ELBO CVAE}
\end{multline}
\noindent where $\text{KL}$ refers to the Kullback–Leibler divergence. Expectations are evaluated in autodifferentiable manner via the reparametrization trick~\cite{kingma_auto-encoding_2022}.

\subsubsection{Conditional prior term: $p(\bz|\by)$.} We design the conditional prior $p(\bz|\by)$ as a bidirectional bottom-up and top-down encoder as per Fig.~\ref{fig: CNVAE model}. Following the strategy of~\cite{vahdat_nvae_2020} to tame the KL terms, we parametrize $p(\bz_l|\bz_{<l},\by)$ relative to the unconditional prior $p(\bz_l|\bz_{<l})$. Let $p(\bz_l^i|\bz_{<l})\triangleq \calN(\mu_i(\bz_{<l}),\sigma_i(\bz_{<l}))$ be a Normal distribution for the $i$th variable in $\bz_l$ under unconditional prior, as output by the top-down branch (dark blue branch in Fig.~\ref{fig: CNVAE model}(a)). We define $p(\bz_l^i|\bz_{<l},\by)\triangleq \calN(\mu_i^y,\sigma_i^y)$, with $\mu_i^y\triangleq \mu_i(\bz_{<l})+\Updelta\mu_i^y(\bz_{<l},\by)$, and $\sigma_i^y=\sigma_i(\bz_{<l})\Updelta\sigma_i^y(\bz_{<l},\by)$, where $\Updelta\mu_i^y(\bz_{<l},\by)$ and $\Updelta\sigma_i^y(\bz_{<l},\by))$ are the mean and standard deviation shifts between the conditional and unconditional priors. With such parametrization, if the unconditional prior moves during optimization, so does the conditional prior.

\subsubsection{Variational posterior term: $q(\bz|\bx,\by)$.} Secondly, we design the variational posterior $q(\bz|\bx,\by)$ as a bidirectional bottom-up and top-down encoder that shares its top-down branch with the conditional prior $p(\bx|\by)$ (dark blue branch in Fig.~\ref{fig: CNVAE model}(b)), in the spirit of the NVAE model~\cite{vahdat_nvae_2020}. Further, we parametrize $q(\bz_l|\bz_{<l},\by,\bx)$ relative to the conditional prior $p(\bz_l|\bz_{<l},\by)$, namely $q(\bz_l^i|\bz_{<l},\by,\bx)=\calN(\mu_i^y+\Updelta\mu_i^x(\bz_{<l},\by,\bx),\sigma_i^y\Updelta\sigma_i^x(\bz_{<l},\by,\bx))$, where $\Updelta\mu_i^x(\bz_{<l},\by,\bx)$ and $\Updelta\sigma_i^x(\bz_{<l},\by,\bx)$ are the mean and standard deviation shifts of the variational posterior w.r.t.~the conditional prior $p(\bx|\by)$. This allows the KL term in $\calL_{\text{CVAE}}(\bx,\by)$ to rewrite simply as Eq.~\ref{eq: KL penalty}:
\begin{equation}
    \text{KL}\left[q(\bz_l^i|\bz_{<l},\bx,\by)||p(\bz_l^i|\bz_{<l},\by)\right] = \frac{1}{2}\left(\frac{{\Updelta\mu_i^x}^2}{{\sigma_i^y}^2}+{\Updelta\sigma_i^x}^2-\log{{\Updelta\sigma_i^x}^2}-1\right)
    \label{eq: KL penalty}
\end{equation}

\subsubsection{Conditional joint distribution term: $p(\bx,\bz|\by)$.} Finally, the generative model $p(\bx,\bz|\by)$ is encoded as a bidirectional bottom-up, top-down decoder that shares its branches with the conditional prior (light/dark blue branches in Fig.~\ref{fig: CNVAE model}(c)). 

Altogether, this leads to a CNVAE design with a y-Encoder for the conditional prior (with $\by$ as input, being either an image to segment or a mask to edit), an x-Encoder for the variational posterior (with $\bx$ as input, being a mask) and a Decoder that generates $\bx$ conditioned on $\by$. In addition, to stabilize the training, we add the following hierarchical KL penalties between conditional and unconditional priors (Eq.~\eqref{eq: final CNVAE loss}).
\begin{equation}
    \calL(\bx,\by)\triangleq \calL_{\text{CVAE}}(\bx,\by)-\sum_{l=1}^L \mathbb{E}_{q(\bz_{<l}|\bx,\by)}\big[\text{KL}\left[p(\bz_l|\bz_{<l},\by)\Vert p(\bz_l|\bz_{<l})\right]\big]
    \label{eq: final CNVAE loss}
\end{equation}

\subsubsection{Image-to-Mask $\texttt{CNVAE}_{\text{Seg}}$ implementation:} In this first scenario, we use our proposed CNVAE framework for direct image segmentation. Specifically, $\by\coloneqq I$ is an input medical image and $\bx$ the ground truth segmentation mask. We refer to this model as $\texttt{CNVAE}_{\text{Seg}}$.

\subsubsection{Mask-to-Mask $\texttt{CNVAE}_{\text{Edit}}$ implementation:} In this second scenario, we use our proposed CNVAE framework for editing segmentation masks output by pre-existing image segmentation models $\{f_m(I)\}_{m=1}^M$ towards anatomically correct masks. In this case, $\by=f_m(I)$ is a mask generated by one of the $f_m$ methods, and $\bx$ the corresponding ground truth. We refer to this model as $\texttt{CNVAE}_{\text{Edit}}$. $\texttt{CNVAE}_{\text{Edit}}$ is trained from pairs of (predicted, ground truth) segmentation masks $(\by\!=\!f_m(I),\bx)$ for a collection of $M$ models spanning different architectures. We test editing capabilities at inference time for in-distribution segmentation methods used for training, and for out-of-distribution segmentation methods not seen at training time.

\section{Experiments \& Results}

\begin{figure}
\centering\includegraphics[width=1.0\linewidth]{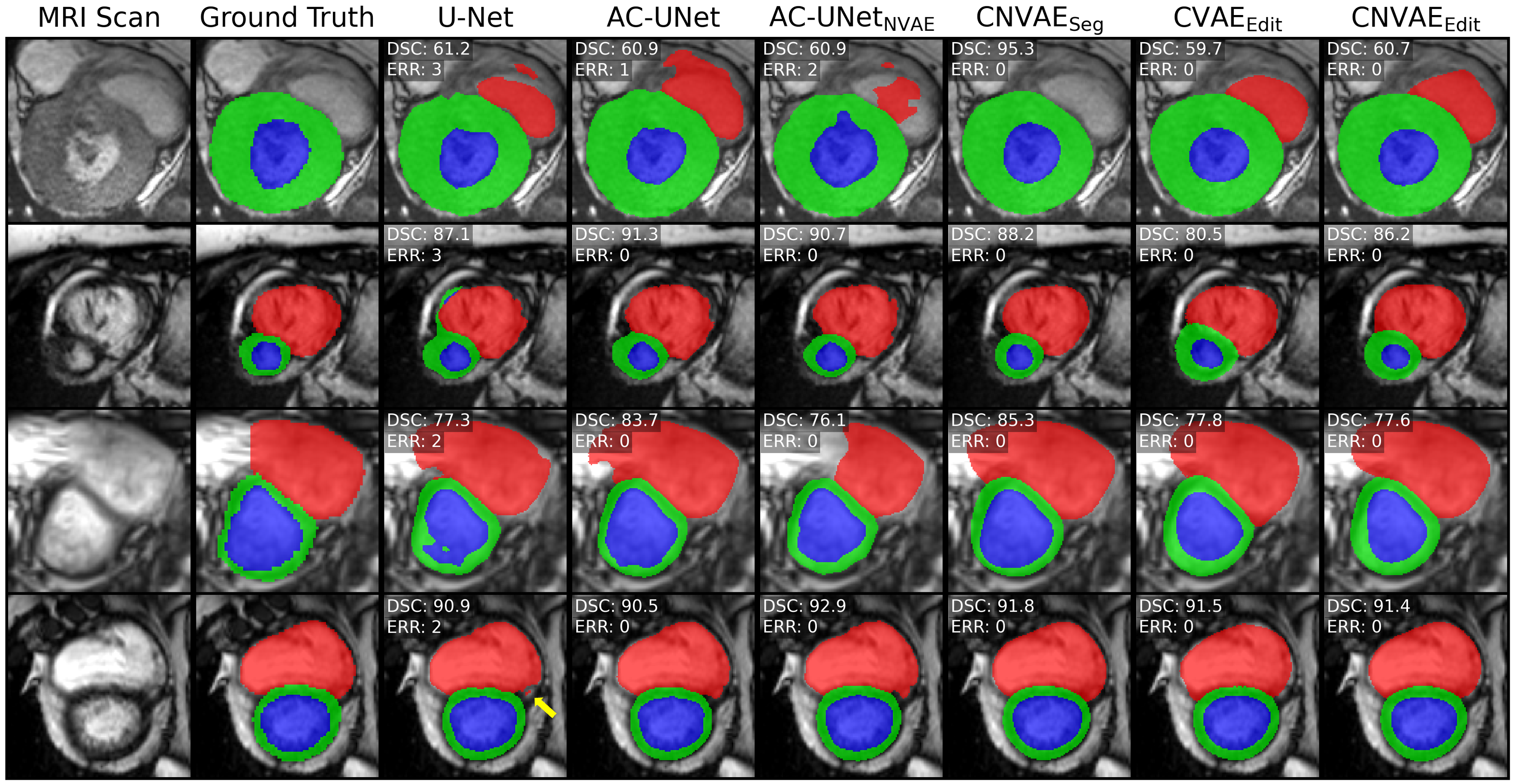}
\caption{Qualitative results from the SOTA and proposed CNVAE models (best viewed with colors, zoomed in). Blue $\equiv$ left ventricle, green $\equiv$ myocardium, red $\equiv$ right ventricle. DSC: Dice score for the segmented 2D slice. ERR: number of violated anatomical criteria. $\texttt{CVAE}_\text{Edit}$ and $\texttt{CNVAE}_\text{Edit}$ (7th and 8th rows) edit the U-Net prediction (3rd row).} \label{fig: example results}
\end{figure}

We demonstrate the \texttt{CNVAE} framework on cardiac MRI segmentation.

\subsubsection{Architecture and Implementation details.} We use a lightweight \texttt{CNVAE} architecture with $L\coloneqq 7$ groups of $D\coloneqq 20$ latent variables with $n_l\coloneqq \{5,4,4,3,3,3,3\}$, and convolutional upsampling and downsampling stems mapping between $H_7\times W_7\coloneqq 16\times 16$ and $H\times W = 128\times 128$. 
The encoder and decoder blocks use the same ResidualCell design as in the original NVAE~\cite{vahdat_nvae_2020}. 

The proposed \texttt{CNVAE} framework is implemented in \texttt{PyTorch} and available (models and configurations) at \url{https://github.com/*/}. For $\texttt{CNVAE}_{\text{Seg}}$, we weight the KL penalties in Eq.~\eqref{eq: ELBO CVAE} by $\beta\coloneqq 3$.  
For $\texttt{CNVAE}_{\text{Edit}}$, we simplify the model by discarding the x-Encoder, effectively forcing the variational posterior to match the conditional prior. We train all models for $100$ epochs with the {\texttt{Adam}} optimizer.
The batch size is set to 8, the learning rate to {$0.001$}. All models are trained on a single GPU (Nvidia Tesla T4 16GB, Tesla A30 12GB or Tesla A16 16GB) in around 50 minutes.

\subsubsection{Data.} The ACDC dataset~\cite{bernard_deep_2018} contains $3D$ short-axis cardiac cine MR images of $150$ subjects, including 
ground truth annotations at End-Systole 
and End-Diastole 
for 
the left ventricle, right ventricle and myocardium. We use the split provided in~\cite{bernard_deep_2018} between training data ($|X_{tr}|=100$) and test data ($|X_{ts}|=50$). $X_{ts}$ is only used for the final evaluation. MRI image slices are intensity-normalized using min-max normalization (using the 1st and 99th percentiles), cropped  
and resized to $128\times 128$ pixels.

\subsubsection{Experimental setup and Evaluation.} 
For training, the slices in the stacks of $X_{tr}$ are {randomly} divided between training data ($90\%$) and validation data ($10\%$). 
For $\texttt{CNVAE}_{\text{Edit}}$, the in-distribution segmentation models used at training time include: (1) \texttt{U-Net}~\cite{ronneberger_olafand_fischer_u-net_2015}; (2) \texttt{AttentionUnet}~\cite{oktay_attention_2018}; (3) \texttt{SwinUNETR}~\cite{hatamizadeh_swin_2022}; and (4) \texttt{ResUnet}~\cite{zhang_road_2018}; as implemented in the \texttt{MONAI} open-source framework 
when available. 
The out-of-distribution segmentation model is \texttt{HighResNet}~\cite{li_compactness_2017}.

To quantify segmentation accuracy, we use the $3D$ Dice similarity coefficient at subject-level segmentations. We report the average Dice over the test set over the segmented structures, {over ED and ES}.
We quantify the anatomical validity \%AV via {9} types of anatomical mistakes adapted from~\cite{painchaud_cardiac_2020}: (1-3) holes in the left ventricle/myocardium/right ventricle; (4-6) more than one connected component for the left ventricle/myocardium/right ventricle; (7) right ventricle disconnected from the myocardium; (8-9) left ventricle connected to the background/right ventricle. A prediction is anatomically valid if none of these criteria is satisfied. \%AV is the proportion of anatomically valid predictions. The \%AV of the ground truth labels is {$99.7\%$}.

\subsubsection{Comparison to other methods.} (1) We train fully-supervised segmentation baselines with a standard pixel-wise loss (cross-entropy): \texttt{U-Net}~\cite{ronneberger_olafand_fischer_u-net_2015}, \texttt{SwinUNETR}~\cite{hatamizadeh_swin_2022}, \texttt{ResUNet}~\cite{zhang_road_2018}, \texttt{AttentionUNet}~\cite{oktay_attention_2018}, {\texttt{HighResNet}}~\cite{li_compactness_2017}. (2) We implement the anatomically constrained U-Net framework of~\cite{oktay_anatomically_2018} (\texttt{AC-UNet}) that complements the pixel-wise loss with a shape loss in the latent space of a VAE. (3) We implement a variant of \texttt{AC-UNet}, where the shape loss uses the latent space of an NVAE instead ($\texttt{AC-UNet}_{\text{NVAE}}$). (4) We implement one of the models proposed in~\cite{painchaud_cardiac_2020}, which edits segmentation predictions via a conditional VAE trained to predict the ground truth mask from incorrect predictions (referred to as $\texttt{CVAE}_{\text{Edit}}$). (5) We train a conditional generation model based on diffusion, MedSegDiff~\cite{wu_medsegdiff_2024}.

\subsubsection{Results.} Fig.~\ref{fig: example results} shows qualitative results. Quantitative results are reported in Tab.~\ref{tab: benchmark}. The 4 baseline segmentation models have anatomical validity rates \%AV in the range $65-84\%$. The proposed segmentation model $\texttt{CNVAE}_\text{Seg}$ has close to perfect anatomical validity ($99.2\%$). This is $11.3\%$ above the next best segmentation method $\texttt{AC-UNet}_{\text{NVAE}}$. The segmentation accuracy (Dice) of $\texttt{CNVAE}_\text{Seg}$ is not statistically significantly worse from that of baseline models ({p-value}: $0.72$). 

The proposed mask editing method, $\texttt{CNVAE}_\text{Edit}$, preserves the original segmentation accuracy while successfully fixing most anatomical mistakes ($\%\text{AV}=98\%$). This is in contrast to the VAE based alternative, $\texttt{CVAE}_\text{Edit}$, that incurs a drop in Dice and lower \%AV. Finally, the approach is robust to out-of-distribution inputs from the \texttt{HighResNet} model. Despite a much degraded original $\%\text{AV}=27.6\%$ of \texttt{HighResNet}, the proposed approach retains a high rate of successful corrections ($\%\text{AV}=89.9\%$).

Both proposed models are also lightweight, comparing favorably to all other models in terms of number of parameters and FLoating Point OPerations (FLOPs). 



\begin{table}[htb]
    \centering
    \begin{tabular}{p{0.35\textwidth} p{0.1\textwidth} p{0.1\textwidth} p{0.2\textwidth} p{0.25\textwidth}}
    \bottomrule
        \textbf{Method} & \textbf{Dice}$\uparrow$ & \textbf{\%AV}$\uparrow$ & \textbf{\#Param (M)}$\downarrow$ & \textbf{GFLOPs/sample}$\downarrow$ \\
    \bottomrule
        \rowcolor{Gray}
        \multicolumn{5}{c}{\textbf{Segmentation methods (image-to-mask)}:} \\
        4 \texttt{Segmentation Baselines} & $\underline{90.8}_{\pm 0.67}$ & $73.6_{\pm 8.0}$ & $6.3\text{ to }31.7$ & $\underline{2.4}\text{ to }24.1$ \\
        \texttt{AC-UNet} & $\mathbf{91.5}$ & $85.3$ & $31.0$ & $24.1$ \\
        $\texttt{AC-UNet}_{\text{NVAE}}$ & $\mathbf{91.5}$ & ${87.9}$ & $31.0$ & $24.1$ \\
        \texttt{MedSegDiff} & $88.2$ & $13.8$ & $522.0$ & $5200$ \\
        \textbf{$\texttt{CNVAE}_{\text{Seg}}$ (Proposed)} & {$90.6$} & $\mathbf{99.2}$ & $\mathbf{3.4}$ & 
        $\mathbf{0.48}$ \\
        \rowcolor{Gray}
        \multicolumn{5}{c}{\textbf{Mask Editing methods (mask-to-mask)}:} \\
        $\texttt{CVAE}_{\text{Edit}}$ & $88.2$ & $92.1$ & +$4.6$ & +$1.74$ \\
        \textbf{$\texttt{CNVAE}_{\text{Edit}}$ (Proposed)} & $\underline{90.8}$ & $\underline{98.0}$ & +${3.4}$
        & +${0.48}$ \\
        \rowcolor{Gray}
        \multicolumn{5}{c}{\textbf{Mask Editing methods, on out-of-distribution inputs from \texttt{HighResNet}}:} \\
        $\texttt{HighResNet}$ & $90.3$ & $27.6$ & $1.1$ & $36.1$ \\
        $\texttt{CVAE}_{\text{Edit}}$ & $87.8$ & $89.0$ & +$4.6$ & +$1.74$ \\
        \textbf{$\texttt{CNVAE}_{\text{Edit}}$ (Proposed)} & ${89.0}$ & ${89.9}$ & +${3.4}$ & +${0.48}$ \\
    \bottomrule
    \end{tabular}
    \caption{Evaluation on the test ACDC dataset. AV: Anatomical Validity. Best results in \textbf{bold} and second best \underline{underlined}. The 4 segmentation baselines are: \texttt{U-Net}, \texttt{SwinUNETR}, \texttt{ResUNet}, \texttt{AttentionUNet}. Mask editing methods add parameters/FLOPs on top of the pre-existing segmentation models, this is marked with a $+$.}
    \label{tab: benchmark}
\end{table}

\begin{figure}
\centering\includegraphics[width=1.0\linewidth]{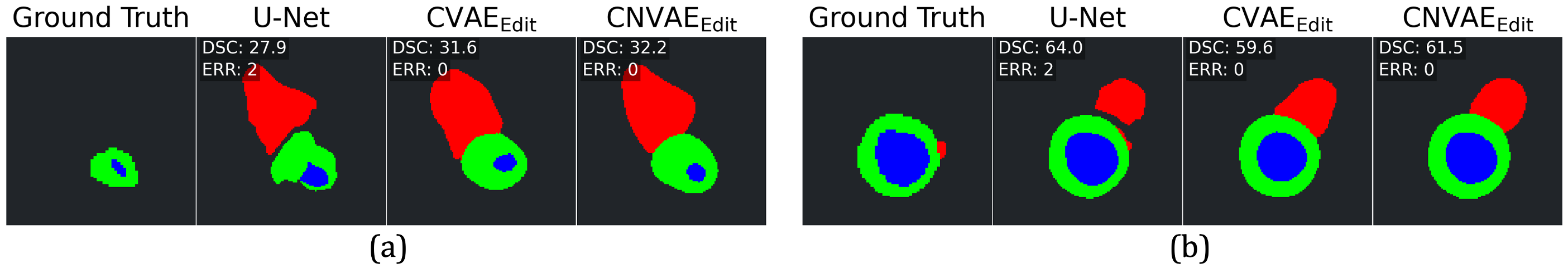}
\caption{Example corrections of the U-Net prediction by two mask editing models: the SOTA $\texttt{CVAE}_\text{Edit}$ and the proposed $\texttt{CNVAE}_\text{Edit}$ (best viewed with colors and zoomed in). Blue $\equiv$ left ventricle, green $\equiv$ myocardium, red $\equiv$ right ventricle. ERR: number of violated anatomical criteria. (a) and (b): two examples.} \label{fig: example corrections}
\end{figure}

\subsubsection{Limitations.} $\texttt{CNVAE}_\text{Edit}$ has no access to the original MRI image when editing the input segmentation. Therefore, although it generally successfully fixes anatomical mistakes, the correction is not guaranteed to be in adequation with the ground truth. Two such examples are provided in Fig.~\ref{fig: example corrections}.

\section{Conclusion} 

This paper introduces a Bayesian conditional generative model, \texttt{CNVAE}. \texttt{CNVAE} extends the \texttt{NVAE} generative framework by allowing to condition on additional image-based inputs. Within this conditional generative paradigm, we address tasks of image segmentation ($\texttt{CNVAE}_\text{Seg}$) or segmentation mask editing ($\texttt{CNVAE}_\text{Edit}$). We demonstrate its effectiveness in overcoming the problem of anatomical correctness of segmentation outputs on an example task of MRI cardiac segmentation, reaching anatomical validity rates of $98-99\%$, $11\%$ above the state-of-the-art among segmentation methods. 
Furthermore, our mask editing method fixes anatomical mistakes without incurring a drop in segmentation accuracy as measured by Dice. Future efforts will focus on extending additional conditioning in \texttt{CNVAE} models (\textit{e.g.}, disease status).

%
%
\bibliographystyle{splncs04}
\bibliography{bibliography, bib}

\end{document}